\documentclass[twocolumn,description,aps]{revtex4-1}
\usepackage{graphicx,epstopdf,amsmath,amssymb,multirow,CJK,color,tabularx}
\usepackage[german,english]{babel}
\usepackage[colorlinks, linkcolor=blue,anchorcolor=blue,citecolor=blue,urlcolor=blue]{hyperref}
\usepackage{multirow}

\begin{document}
\title{Symmetry-protected coexistence of a nodal surface and multiple types of Weyl fermions in $P6_3$-$\text{B}_{30}$}

\author{Xiao-Jing Gao$^{1}$}
\author{Yanfeng Ge$^{1}$}
\author{Yan Gao$^{1}$}\email{yangao9419@ysu.edu.cn}

\affiliation{$^{1}$State Key Laboratory of Metastable Materials Science and Technology $\&$ Hebei Key Laboratory of Microstructural Material Physics, School of Science, Yanshan University, Qinhuangdao 066004, China}

\date{\today}

\begin{abstract}
The coexistence of topological states with different dimensionalities in a single crystalline system offers a unique platform to study the interplay of distinct fermionic excitations. Here, integrating first-principles calculations with symmetry analysis, we propose the three-dimensional boron allotrope $P6_3$-$\text{B}_{30}$ as an ideal, structurally stable candidate for exploring multidimensional topological physics. Benefiting from the practically negligible spin-orbit coupling of the light-element framework, $P6_3$-$\text{B}_{30}$ operates as a pristine spinless topological semimetal. We show that the combined time-reversal and twofold screw symmetry ($\mathcal{T}S_{2z}$) enforces a robust two-dimensional nodal surface on the $k_z = \pi$ plane via a Kramers-like degeneracy. Concurrently, the system hosts a diverse set of zero-dimensional Weyl fermions---including an unconventional double-Weyl point ($\mathcal{C} = -2$), conventional Type-I WPs ($\mathcal{C} = -1$), and completely tilted Type-II WPs ($\mathcal{C} = +1$)---emerging at the high-symmetry points $\Gamma$ and K, as well as along the H–K path, protected by $C_6$ and $C_3$ crystalline rotational symmetries. Crucially, the substantial momentum-space separation between the nodal surface and Weyl points allows for their unambiguous independent resolution. Calculations of the (100) surface states reveal distinct, nontrivial Fermi arcs connecting Weyl nodes of opposite chirality. This work establishes $P6_3$-$\text{B}_{30}$ as a compelling material platform for investigating the physics of multidimensional hybrid topological fermions and their interplay.
\end{abstract}

\date{\today} \maketitle

\section{INTRODUCTION}\label{sec_introduction}

Since the foundational discovery of topological insulators~\cite{ref1,ref2}, the classification of topological phases has undergone rapid expansion, encompassing topological crystalline insulators~\cite{ref3,ref4,ref5}, topological superconductors~\cite{ref6,ref7,ref8}, and topological semimetals (TSMs)~\cite{ref9,ref10,ref11,ref12}. Among these, TSMs have garnered immense theoretical and experimental interest due to their unconventional electronic structures and associated anomalous transport responses.  Depending on the dimensionality of the topological or symmetry-protected band crossings, TSMs are generally classified into zero-dimensional (0D) nodal points (e.g., Weyl and Dirac semimetals)~\cite{ref13,ref14,ref15,ref16,ref17,ref18}, one-dimensional (1D) nodal lines (NLs)~\cite{ref19,ref20,ref21}, and two-dimensional (2D) nodal surfaces (NSs)~\cite{ref22,ref23,ref24,ref25}.

Despite significant progress in cataloguing TSMs that host a single class of band degeneracy, materials that simultaneously realize topological objects of different dimensionalities, such as coexisting Weyl points (WPs) and NSs within the same electronic structure, remain comparatively rare and largely underexplored~\cite{ref26,ref27,ref28}. Such multidimensional hybrid TSMs provide a unique platform for investigating the interplay between topological objects of different dimensionalities, including their mutual constraints, connectivity of surface states, and possible interference in transport phenomena~\cite{ref29,ref30}. Crucially, for these hybrid states to be experimentally tractable, the distinct topological objects must be well separated in momentum and energy space to prevent mutual interference from obscuring their individual signatures. Identifying concrete, stable materials that satisfy these stringent criteria is therefore of considerable importance.

\begin{figure*}
    \centering
    \includegraphics[width=1\linewidth]{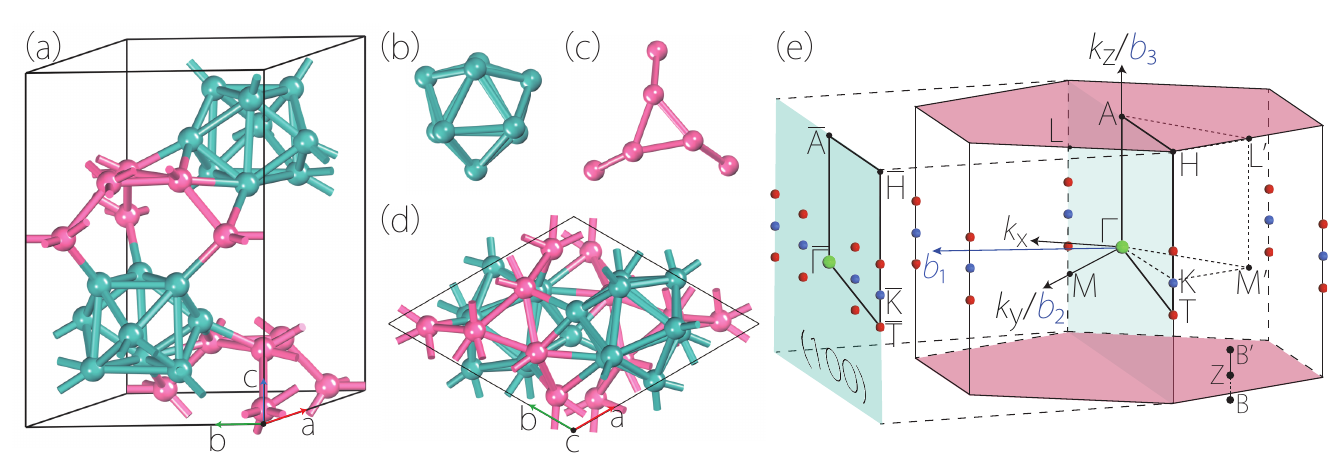}
    \caption{(Color online) Crystal structure and Brillouin zone (BZ) of $P6_3$-B$_{30}$. (a) The primitive cell of $P6_3$-$\text{B}_{30}$ showing the 30-atom hexagonal lattice. (b) Top views of the cage-like B$_9$ unit (c) and the bilayer triangular B$_6$ unit. (d) Top view of the $P6_3$-B$_{30}$ primitive cell. (e) The first BZ and its projection onto the (100) surface. The red, blue, and green dots illustrate Weyl points with topological charges of +1, $-$1, and $-$2, respectively. The pink plane represents the nodal surface located at $k_z = \pi$, while the green plane represents the (100) surface projection.}
    \label{fig:1}
\end{figure*}

Currently, existing proposals for the coexistence of WP and NS states heavily rely on transition-metal or heavy-element compounds (e.g., Pd$_4$$X$ ($X$ = S, Se)~\cite{ref31} and NiAsS~\cite{ref32}, CoSi\cite{ref33}). In such materials, strong spin-orbit coupling (SOC) invariably induces large band splittings, frequently opening gaps at the nodal crossings and driving the system into trivial or insulating phases. To fully capture the intrinsic topology of coexisting semimetallic states, light-element systems (such as carbon or boron) are highly advantageous. The negligible SOC in these materials permits the electrons to be treated effectively as spinless fermions, thereby preserving the structural integrity of symmetry-enforced nodal crossings. Boron, with its versatile multicenter bonding configurations~\cite{ref34,ref35,ref36,ref37}, has emerged as a fertile playground for such spinless topological phases, with 3D allotropes frequently hosting isolated NLs and NSs~\cite{ref38,ref39,ref40,ref41,ref42,ref43,ref44,ref45}. However, a 3D boron material hosting the clean coexistence of isolated WPs and NSs has remained elusive.

In this work, we propose a previously unreported 3D boron allotrope, $P6_3$-$\text{B}_{30}$, and systematically investigate its structural stability and electronic properties using first-principles calculations combined with symmetry analysis. We demonstrate that $P6_3$-$\text{B}_{30}$ hosts a rare coexistence of a 2D NS and multiple types of 0D WPs. Protected by $C_3$ and $C_6$ crystalline rotational symmetries, these 0D nodes comprise a rich topological catalog: conventional Type-I WPs ($\mathcal{C} = -1$) at the K and K$'$ points, completely tilted Type-II WPs ($\mathcal{C} = +1$) on the H–K path, and an unconventional double-Weyl point ($\mathcal{C} = -2$) at the $\Gamma$ point. Concurrently, a robust NS is pinned to the $k_z = \pi$ plane, enforced by the combination of time-reversal ($\mathcal{T}$) and twofold screw ($S_{2z}$) symmetries. The nontrivial topology is further corroborated by the emergence of distinct Fermi arcs on the (100) surface. Crucially, these 0D and 2D topological states are fundamentally decoupled in momentum space. This wide momentum-space separation not only provides an ideal setting for the independent experimental resolution of their respective surface signatures but also establishes $P6_3$-$\text{B}_{30}$ as a pristine candidate for exploring the physical interactions within multidimensional hybrid topological systems.

\begin{table*}[!th]
	\renewcommand{\thetable}{I} 
	\caption{\label{tab:I} Structural parameters, mechanical properties, and topological characteristics of $P6_3$-B$_{30}$ and selected boron allotropes, including space groups, lattice parameters, bond lengths($d_{\text{B-B}}$), density ($\rho$), bulk modulus ($B$), shear modulus ($G$), total energy per atom ($E_{\text{tot}}$), topological properties.}
	\centering
	\begin{tabular*}{2.0\columnwidth}{@{\extracolsep{\fill}}ccccccccccc}
		\hline
        \hline
		\multirow{2}{*}{Structure} & \multirow{2}{*}{Space group} & \multicolumn{1}{c}{a} & \multicolumn{1}{c}{b} & \multicolumn{1}{c}{c} & \multicolumn{1}{c}{$d_{\text{B-B}}$} & \multicolumn{1}{c}{$\rho$} & \multicolumn{1}{c}{B} & \multicolumn{1}{c}{G} & \multicolumn{1}{c}{$E_{\text{tot}}$}  & \multirow{2}{*}{Properties}  \\
		& & (\AA) & (\AA) & (\AA) & (\AA) & ($g/cm^{-3}$) & (GPa) & (GPa) & (eV/atom) &  \\ 
		\hline
    $P6_3$-B$_{30}$  & $P6_3$  & 5.65  & 5.65  & 8.50  & 1.67–2.06    & 2.29  & 161.65    & 71.21     & -6.37  & NP-NSSM     \\
    H-boron       & $P6_3/mmc$   & 6.06  & 6.06  & 9.91 & 1.62–1.71   & 0.91  & 70.42     & 48.06     & -5.84  & NLSM        \\
    Ort-B$_{32}$  & $Cmcm$       & 6.61  & 8.09  & 4.98 & 1.63–2.09   & 2.16  & 194.00    & 179.24    & -6.29  & NLSM        \\
    3D borophene  & $C2/m$       & 5.47  & 2.81  & 1.81 & 1.73–1.89   & 2.52  & 242.34    & 406.04    & -6.31  & NLSM        \\
    Porous-B$_{18}$ & $P6_3/m$   & 6.72  & 6.72  & 3.85 & 1.73–2.02   & 2.14  & 168.91    & 101.76    & -6.34  & NLSM        \\
    3D $\alpha'$-boron & $Cmcm$   & 7.73   & 8.23  & 5.07  & 1.66–1.85  & 1.78   & 166.16    & 116.73    & -6.36  & NLSM     \\
    AB-16-Pnnm    & $Pnnm$       & 3.20  & 8.48  & 4.50   & 1.61–1.91   & 2.35   & 218.32    & 144.27    & -6.42  & NLSM     \\
    CT-B$_{24}$   & $P6_3mc$     & 4.99  & 4.99  & 8.39  & 1.66–2.02  & 2.37  & 172.63    & 147.93    & -6.47  & NSSM        \\
    Pnma-B$_{60}$ & $Pnma$       & 11.82 & 4.90  & 7.52  & 1.72–1.87  & 2.47  & 235.76    & 249.52    & -6.60  & NL-NSSM     \\
    $\gamma$-B$_{28}$ & $Pnnm$   & 5.04  & 5.61  & 6.92  & 1.66–1.90  & 2.57  & 245.22    & 281.91    & -6.68  & Insulator   \\
    $\alpha$-B$_{12}$& $R\overline{3}m$  & 4.89  & 4.89  & 12.55   & 1.67–1.80  & 2.48   & 238.11  & 267.54  & -6.70   & Insulator  \\
		\hline\hline
    \label{table:1}
	\end{tabular*}
\end{table*}

\begin{figure}
    \centering
    \includegraphics[width=0.98\linewidth]{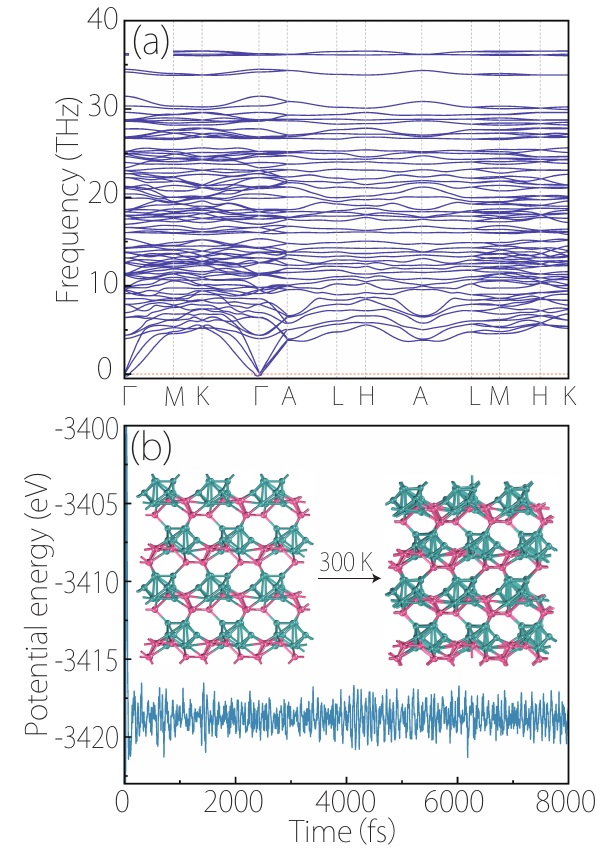}
    \caption{(Color online) Stability characterization of $P6_3$-B$_{30}$. (a) Phonon dispersion spectrum along high-symmetry directions calculated using a $3 \times 3 \times 1$ supercell. (b) Total potential energy evolution from AIMD simulation at 300 K. The left and right insets show the atomic configurations before and after 8 ps of simulation, respectively.}
    \label{fig:2}
\end{figure}

\section{METHOD}\label{sec_method}

Our first-principles calculations were performed within the framework of density functional theory (DFT) as implemented in the VASP code~\cite{ref46}. The electron-ion interactions were described by the projector augmented-wave (PAW) method~\cite{ref47}, and the exchange-correlation functional was treated using the generalized gradient approximation (GGA) parameterized by Perdew, Burke, and Ernzerhof (PBE)~\cite{ref48}. Consistent with our treatment of $P6_3$-$\text{B}_{30}$ as a spinless fermion system, spin-orbit coupling (SOC) was not included in the primary calculations. The plane-wave cutoff energy was set to 500 eV. Structural relaxations were performed until the total energy and residual forces converged to $10^{-6}$ eV and $10^{-3}$ eV/\AA, respectively. The dynamical stability was assessed by calculating the phonon dispersion curves via the finite displacement method using a $3 \times 3 \times 1$ supercell, as implemented in the Phonopy package~\cite{ref49}. To evaluate thermodynamic stability, \textit{ab initio} molecular dynamics (AIMD) simulations were carried out at 300 K utilizing an NVT ensemble with a Nos\'e-Hoover thermostat~\cite{ref50} for a duration of 8 ps. To extract the topological properties, we constructed a maximally localized Wannier function (MLWF) tight-binding Hamiltonian using the Wannier90 package~\cite{ref51} by projecting the DFT Bloch states onto the boron $s$ and $p$ atomic orbitals. The surface state spectra and Fermi arcs were subsequently computed via the iterative surface Green's function method using WannierTools~\cite{ref52}.

\section{RESULTS}\label{sec_results}

\subsection{Crystal Structure and Stability}
\label{subsec:crystal_stability}

$P6_3$-$\text{B}_{30}$ crystallizes in the hexagonal space group $P6_3$ (No. 173). The conventional unit cell comprises 30 boron atoms, yielding optimized lattice parameters of $a = b = 5.65$~\AA~and $c = 8.50$~\AA. As illustrated in Figs.~\ref{fig:1}(a) and \ref{fig:1}(d), the structural framework is composed of two primary building blocks: cage-like $\text{B}_9$ units [Fig.~\ref{fig:1}(b)] and bilayer triangular $\text{B}_6$ units [Fig.~\ref{fig:1}(c)]. These structural motifs intertwine to form a layered network, stacked along the $k_z$ direction and related by a sixfold screw rotation symmetry. The B--B bond lengths vary from 1.67~\AA~to 2.06~\AA, and the calculated mass density is $\rho = 2.29$ g/cm$^3$. The total energy per atom for $P6_3$-$\text{B}_{30}$ is $E = -6.37$ eV/atom; as detailed in Table~\ref{tab:I}, this lies within a highly competitive energetic window relative to other synthesized and theoretically proposed boron allotropes, suggesting experimental feasibility.

We systematically verified the stability of $P6_3$-$\text{B}_{30}$ from dynamic, thermodynamic, and mechanical perspectives. As shown in Fig.~\ref{fig:2}(a), the phonon dispersion along all high-symmetry paths is devoid of intrinsic imaginary frequencies. The negligible imaginary acoustic modes ($<$0.2 THz) near the $\Gamma$ point are standard numerical artifacts originating from the finite supercell size and grid interpolation, which slightly break the acoustic sum rule, and do not reflect any structural instability. Thermodynamic stability was confirmed via AIMD simulations on a $2 \times 2 \times 2$ supercell at 300 K with a 1-fs time step. As depicted in Fig.~\ref{fig:2}(b), the total potential energy exhibits minor thermal fluctuations around a stable mean value, and the atomic configurations remain fully intact after 8 ps. Finally, the mechanical stability was validated by computing the elastic stiffness constants $C_{ij}$. The obtained values, $C_{11} = 430.35$ GPa, $C_{12} = 69.97$ GPa, $C_{13} = 74.30$ GPa, $C_{33} = 221.61$ GPa, and $C_{44} = 19.79$ GPa, rigorously satisfy the Born stability criteria for hexagonal lattices~\cite{ref53}: (i) $C_{11} > |C_{12}|$, (ii) $2C_{13}^2 < C_{33}(C_{11} + C_{12})$, and (iii) $C_{44} > 0$. Together, these results provide robust evidence for the stability of the $P6_3$-$\text{B}_{30}$ phase.

\begin{figure*}[!th]
    \centering
    \includegraphics[width=1\linewidth]{figure_3.pdf}
    \caption{(Color online) Electronic band structure and topological properties of $P6_3$-B$_{30}$. (a) Band structure of $P6_3$-B$_{30}$ along high-symmetry paths. The red and cyan bands represent the highest occupied band and the lowest unoccupied band, respectively. The left inset shows the band dispersion along the direction normal to the NS, with B-Z-B$^\prime$ indicating a generic path. The right inset provides a magnified view of the band crossings along the H-K path. (b) Projected density of states (PDOS) of $P6_3$-B$_{30}$. (c) Real-space charge density at a generic momentum point P. (d) Gap mapping in the $k_y$--$k_z$ plane, showing the nodal surface (NS) and Weyl points ($W_1, W_2, W_3$). (e) Schematic illustration of generic paths on the $k_z = \pi$ plane. (f--g) Band structures along generic paths within the $k_z = \pi$ plane. (h) Wannier charge centers for the three WPs W$_1$, W$_2$, and W$_3$. (i) Three dimensional band structure of the W$_2$ point in the $k_x$--$k_y$ plane.}
    \label{fig:3}
\end{figure*}

\subsection{Electronic Properties and Band Topology}

The intrinsic topological properties of $P6_3$-$\text{B}_{30}$ originate from its bulk electronic structure. Fig.~\ref{fig:3}(a) presents the calculated spinless band structure along high-symmetry lines. The fundamental features are dictated by the crossings between the highest occupied band (index 45, colored red) and the lowest unoccupied band (index 46, colored cyan) near the Fermi level ($E_F$). A projected density of states (PDOS) analysis reveals that the low-energy states near $E_F$ are overwhelmingly dominated by the $p$ orbitals of the bilayer triangular $\text{B}_6$ structural units. This orbital origin aligns well with the calculated real-space charge density at a generic momentum point P, as illustrated in Fig.~\ref{fig:3}(c).

The band crossings between the 45th and 46th bands give rise to two distinct, spatially separated topological features in momentum space. The first class consists of zero-dimensional (0D) Weyl points located at the high-symmetry points K and $\Gamma$, as well as along the H--K path, denoted as $W_1$, $W_2$, and $W_3$, respectively. The second class is a two-dimensional (2D) topological nodal surface (NS) rigidly pinned to the $k_z = \pi$ plane (the A--L--H plane). The 45th and 46th bands exclusively form this NS with a remarkably narrow energy variation ($\Delta E \approx 0.184$ eV) and no parasitic crossing from other trivial bands, establishing an exceptionally clean background for experimental detection.

\begin{table*}[!th]
	\renewcommand{\thetable}{II} 
	\caption{\label{tab:II} Calculated Weyl points in $P6_3$-B$_{30}$: momentum coordinates, energies, and topological charges.}
	\centering
	\begin{tabular*}{2.0\columnwidth}{@{\extracolsep{\fill}}ccccc}
		\hline
        \hline
		\multicolumn{1}{c}{Nodes} & \multicolumn{1}{c}{Momentum coordinates} & \multicolumn{1}{c}{Energy(eV)} & \multicolumn{1}{c}{Chirality} & \multicolumn{1}{c}{Number} \\
		\hline
        \multirow{4}{*}{W$_{3}$} & \multicolumn{1}{c}{(-0.333, 0.667, 0.145)}  & \multirow{4}{*}{-0.343}  & \multirow{4}{*}{+1}  & \multirow{4}{*}{4}           \\
         &  \multicolumn{1}{c}{(-0.333, -0.333, -0.145)} &  &  \\ 
         &  \multicolumn{1}{c}{(-0.667, 0.333, 0.145)} &  &  \\ 
         &  \multicolumn{1}{c}{(-0.667, 0.333, -0.145)} &  &  \\    
        \multirow{1}{*}{W$_{2}$} & (0, 0, 0)  & 0.322       & -2      & 1           \\
        \multirow{2}{*}{W$_{1}$} & (0.333, 0.333, 0)  & \multirow{2}{*}{-0.447}  & \multirow{2}{*}{-1}  & \multirow{2}{*}{2}      \\
         &  \multicolumn{1}{c}{(-0,333, -0,333, 0)} &  &  \\            
		\hline\hline
        \label{table:2}
	\end{tabular*}
\end{table*}

\subsection{Symmetry-Enforced Two-Dimensional Nodal Surface}

We first address the mathematical and physical origin of the 2D NS. To verify the omnipresent degeneracy across the $k_z = \pi$ plane, we calculated the band dispersion along several arbitrarily chosen in-plane paths, labeled $\text{C}_n$ and $\text{D}_n$ ($n=1\text{--}2$) in Fig.~\ref{fig:3}(e). As depicted in Figs.~\ref{fig:3}(f) and \ref{fig:3}(g), the 45th and 46th bands remain perfectly degenerate everywhere within this plane. Crucially, along paths normal to the NS (e.g., along $k_z$), the bands split linearly [see the inset of Fig.~\ref{fig:3}(a)], which is the hallmark dispersion signature of NS fermions.

The formation of this NS is not accidental; rather, it is strictly enforced by the non-symmorphic symmetries of the $P6_3$ space group. In our spinless framework, the system respects time-reversal symmetry ($\mathcal{T}$) and a twofold screw rotation along the $c$-axis, defined as $S_{2z} = \{C_{2z} | 0 0 \frac{1}{2}\}$. When applied to a Bloch state, the combined antiunitary operator $\mathcal{T}S_{2z}$ satisfies the relation:
\begin{equation}
	(\mathcal{T}S_{2z})^2 = \mathcal{T}^2 \{C_{2z} | 0 0 \textstyle{\frac{1}{2}}\}^2 = e^{-i k_z}.
\end{equation}
Evaluated strictly on the BZ boundary plane at $k_z = \pi$, this relation yields $(\mathcal{T}S_{2z})^2 = e^{-i\pi} = -1$. This algebra precisely mimics the behavior of the time-reversal operator in spin-$\frac{1}{2}$ systems. As a result, every generic momentum point \textit{$\mathbf{k}$} on the $k_z = \pi$ plane inherently hosts a twofold degenerate band crossing, culminating in a robust, symmetry-protected Nodal Surface.

\subsection{Symmetry and Topology of the Zero-Dimensional Weyl Points}

Beyond the 2D nodal surface, we systematically scanned the bulk BZ to investigate the zero-dimensional (0D) topological nodes. We identified a total of seven isolated Weyl points (WPs) in the first BZ. Group-theoretical analysis of the band structure reveals that at the high-symmetry points, the crossing bands correspond to 2D irreducible representations (irreps)—specifically, K$_2$+K$_3$ at the K point and $\Gamma_3$+$\Gamma_5$ at the $\Gamma$ point. However, at generic momentum points in their immediate vicinity, the spatial symmetry is lowered, and the bands split into 1D irreps. This symmetry-breaking dictates that these degeneracies are strictly isolated 0D WPs rather than continuous nodal lines. 

To rigorously characterize the topological nature of these nodes, we calculated the intrinsic Chern number ($\mathcal{C}$) for each WP by integrating the Berry curvature over a closed sphere enclosing each node [Fig.~\ref{fig:3}(h)]. The WPs can be classified into three distinct sets based on their topological charges and momentum locations (detailed in Table~\ref{tab:II}):
(i) Four $W_3$ points on the H–K path carrying $\mathcal{C} = +1$,
(ii) One $W_2$ point at the $\Gamma$ point carrying an unconventional charge of $\mathcal{C} = -2$,
(iii) Two $W_1$ points at the K and K$'$ points carrying $\mathcal{C} = -1$.
Remarkably, the algebraic sum of the topological charges over the entire BZ is exactly zero $[(+1 \times 4) + (-2 \times 1) + (-1 \times 2) = 0]$, perfectly satisfying the Nielsen-Ninomiya no-go theorem. 

\begin{figure}
    \centering
    \includegraphics[width=0.90\linewidth]{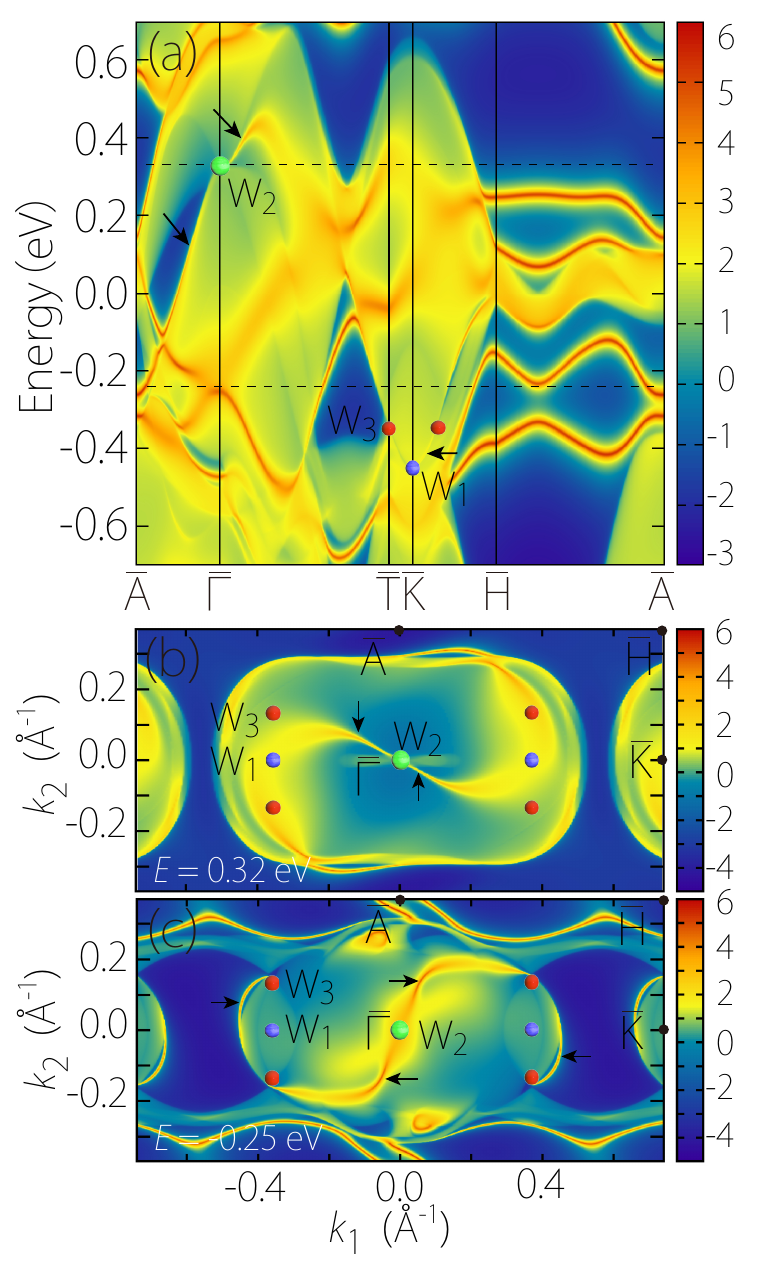}
    \caption{(Color online) Surface states and Fermi arcs. (a) Surface density of states projected onto the (100) surface. (b--c) Constant energy contours at $E = 0.32$ eV and $E = -0.25$ eV, respectively. The red, blue, and green dots indicate the surface projections of the bulk Weyl points. Arrows indicate the flow of the Fermi arcs from positive to negative topological charges.}
    \label{fig:4}
\end{figure}

The diverse topological charges and momentum locations mandate distinct energy dispersion characteristics. The four $W_1$ points exhibit conventional linear band dispersion with opposite slopes in all three momentum directions, which is the hallmark of Type-I Weyl fermions. Interestingly, as shown in the inset of Fig.~\ref{fig:3}(a), the bands crossing at the $W_3$ points exhibit the same sign of slopes along a specific direction, indicating a completely tilted dispersion characteristic of Type-II Weyl fermions. The coexistence of Type-I and Type-II WPs introduces an additional degree of freedom in the topological band dispersion type. In stark contrast to both, the $W_2$ point exhibits a linear dispersion strictly along the out-of-plane $k_z$ direction (the $\Gamma$--A path) but a quadratic dispersion in the in-plane ($k_x, k_y$) directions [Fig.~\ref{fig:3}(i)]. This quadratic in-plane profile is the defining physical signature of an unconventional double-Weyl point, aligning perfectly with its higher-order topological charge of $\mathcal{C} = -2$. 

The stability of these WPs is fundamentally protected by the crystalline rotational symmetries of the $P6_3$ space group. Specifically, $W_1$ is protected by the $C_3$ rotational symmetry. The double-Weyl point $W_2$ is safeguarded by the combined $C_6$ and $\mathcal{T}$ symmetries at the BZ center. Furthermore, for $W_3$, an analysis of the wavefunctions along the high-symmetry path reveals that the two crossing bands possess distinct $C_3$ rotation eigenvalues of $e^{-2\pi i/3}$ and $e^{2\pi i/3}$. Because their eigenvalue ratio is $e^{\pm 2\pi i/3}$, hybridization between these bands is strictly forbidden by symmetry, rendering the $W_3$ band inversion topological and robust.

\subsection{Bulk-Boundary Correspondence and Fermi Arcs}

The most direct experimental hallmark of topological Weyl semimetals is the emergence of nontrivial surface states and Fermi arcs connecting the surface projections of bulk WPs. To verify this bulk-boundary correspondence in $P6_3$-B$_{30}$, we calculated the local density of states on the semi-infinite (100) projected surface. 

Figs.~\ref{fig:4}(a)--\ref{fig:4}(c) map the surface states and the constant energy contours at $E = 0.32$ eV and $E = -0.25$ eV. The surface projections of the WPs and their corresponding topological charges are clearly distinguishable. A rigorous inspection of the Fermi arcs unequivocally corroborates our bulk-WPs Chern number calculations, adhering strictly to the principle that Fermi arcs must originate from nodes with a positive chiral charge and terminate at nodes with a negative chiral charge. As shown in Fig.~\ref{fig:4}(b), exactly two distinct Fermi arcs terminate at the central projected $\Gamma$ point. This double-arc connectivity is the macroscopic manifestation of the $\mathcal{C} = -2$ bulk topological charge of the $W_2$ double-Weyl point. Tracking these arcs outward [Fig.~\ref{fig:4}(c)], these two arcs originate from two adjacent $W_3$ points ($\mathcal{C} = +1$ each). Simultaneously, the remaining $W_3$ points emit individual arcs that terminate at the adjacent $W_1$ points ($\mathcal{C} = -1$ each) [Fig.~\ref{fig:4}(c)]. This intricate, closed-loop surface state topology not only confirms the chiral nature of the coexisting Type-I, Type-II, and double-Weyl points but also provides a distinct, clean signature for future angle-resolved photoemission spectroscopy (ARPES) measurements, completely free from interference by the momentum-separated Nodal Surface.

\section{DISCUSSION AND SUMMARY}\label{sec_discussion}

The search for multidimensional topological phases, where multiple distinct topological band degeneracies coexist within a single Brillouin zone, represents a significant frontier in condensed matter physics. However, most contemporary candidates, such as the $\text{LaSb}_2$ system~\cite{ref26} and heavy-element stannides or silicides~\cite{ref31,ref33}, suffer from two critical limitations: strong momentum-space entanglement between topological features and the deleterious effects of strong spin-orbit coupling, which inevitably gaps out the symmetry-enforced nodal structures.

In contrast, our work identifies $P6_3$-$\text{B}_{30}$ as an exceptional material platform that circumvents these barriers. By leveraging the light-element boron framework, we demonstrate a spinless topological regime where the structural integrity of band crossings is preserved. Most importantly, when compared to existing topological semimetals, $P6_3$-$\text{B}_{30}$ exhibits a significantly richer topological diversity. While prior literature has documented the coexistence of nodal lines with Weyl points~\cite{ref54,ref55}, or standard nodal surfaces with $\mathcal{C} = \pm 1$ fermions~\cite{ref56,ref57}, the simultaneous realization of a 2D nodal surface and higher-order topological fermions, specifically the unconventional $\mathcal{C} = -2$ double-Weyl point, remains elusive. The presence of this higher-order charge, alongside the coexistence of Type-I and tilted Type-II Weyl fermions, positions $P6_3$-$\text{B}_{30}$ as a uniquely complex laboratory of Weyl fermions that remains unmatched by previously reported nodal surface or node-coexisting materials.

The spatial decoupling of these features in momentum space is a key competitive advantage. In many previously reported systems, Weyl points are embedded within nodal planes~\cite{ref26}, leading to significant hybridization and the obscuration of individual surface states. In $P6_3$-$\text{B}_{30}$, the nodal surface is strictly pinned to the $k_z = \pi$ plane, while the various Weyl nodes are well-isolated in the $k_z \neq \pi$ sectors. This decoupling ensures that the resulting surface Fermi arcs, particularly the distinctive double-arc connectivity originating from the $\mathcal{C} = -2$ node, can be resolved independently via ARPES without interference from the nodal surface states.

In summary, based on first-principles calculations, we have systematically investigated the electronic structure and topological properties of $P6_3$-B$_{30}$, revealing the coexistence of two distinct topological states of different dimensions within a single material. Stability analyses confirm that the structure possesses good intrinsic stability, indicating high feasibility for experimental synthesis. Regarding the electronic band structure, our calculations demonstrate the simultaneous presence of a 2D NS and 0D WPs, both of which are topologically protected by symmetries as confirmed by further symmetry analysis. Furthermore, the nontrivial surface states connecting WPs of opposite chirality on the (100) surface not only confirm the topological charges carried by the WPs but also provide clear signatures for future experimental detection of these topological states. Overall, $P6_3$-B$_{30}$ enriches the family of topological boron allotropes and serves as an ideal platform for investigating the interplay between nodal surfaces and nodal points.

\begin{acknowledgments}

This work was supported by the National Natural Science Foundation of China (Grants No.~12304202), Hebei Natural Science Foundation (Grant No.~A2023203007), Science Research Project of Hebei Education Department (Grant No.~BJK2024085), and Cultivation Project for Basic Research and Innovation of Yanshan University (No.~2022LGZD001).

\end{acknowledgments}

\end{document}